\documentclass[10pt,conference,compsocconf]{IEEEtran}
\usepackage{times}
\usepackage{cite}
\usepackage{booktabs}                          

\let\labelindent\relax
\usepackage{enumitem}
\usepackage[small]{caption}
\usepackage{subfigure}
\usepackage{xcolor}
\usepackage[hyphens]{url}
\usepackage{url}
\usepackage{graphicx}
\newcommand{\descr}[1]{\smallskip\noindent\emph{#1}}

\let\OLDthebibliography\thebibliography
\renewcommand\thebibliography[1]{
  \OLDthebibliography{#1}
  \setlength{\parskip}{0pt}
  \setlength{\itemsep}{1pt plus 0.2ex}
}

\begin{document}
\pagenumbering{gobble}
%
\title{\textbf{\Large A Privacy-Preserving Architecture for the Protection of Adolescents \\[-1.5ex] in Online Social Networks}\\[-0.9ex]}

\author{~\\[-0.4ex]\large Markos Charalambous$^{1}$, Petros Papagiannis$^{1}$, Antonis Papasavva$^{1}$, Pantelitsa Leonidou$^{1}$,\\ Rafael Constaninou$^{2}$, Lia Terzidou$^{3}$, Theodoros Christophides$^{1}$, Pantelis Nicolaou$^{2}$ \\Orfeas Theofanis$^{4}$, George Kalatzantonakis$^{4}$, Michael Sirivianos$^{1}$\\[0.3ex]

$^{1}$Cyprus University of Technology, Limassol Cyprus\\
$^{2}$Cyprus Research and Innovation Center, Nicosia, Cyprus \\ $^{3}$Aristotle University of Thessaloniki, Thessaloniki, Greece\\
$^{4}$LSTech LTD, Milton Keynes, United Kingdom\\

Email: \{marcos.charalambous, petros.papagiannis, t.christophides, michael.sirivianos\}@cut.ac.cy,\\ \{as.papasavva, pl.leonidou\}@edu.cut.ac.cy, \{r.constantinou, p.nicolaou\}@cyric.eu, lterz@csd.auth.gr,\\ 
\{orfetheo, george\}@lstech.io 
\\[1.0ex]}

\maketitle

\begin{abstract}
Online Social Networks (OSN) constitute an integral part of people's every day social activity. 
Specifically, mainstream OSNs, such as Twitter, YouTube, and Facebook are especially prominent in adolescents' lives for communicating with other people online, expressing and entertain themselves, and finding information. 
However, adolescents face a significant number of threats when using online platforms.
Some of these threats include aggressive behavior and cyberbullying, sexual grooming, false news and fake activity, radicalization, and exposure of personal information and sensitive content. 
There is a pressing need for parental control tools and Internet content filtering techniques to protect the vulnerable groups that use online
platforms. 
Existing parental control tools occasionally violate the privacy of
adolescents, leading them to use other communication channels to
avoid moderation.
In this work, we design and implement a user-centric Cybersafety Family Advice Suite (CFAS) with Guardian Avatars aiming at preserving the privacy of the individuals towards their custodians and towards the advice tool itself. 
Moreover, we present a systematic process for designing
and developing state of the art techniques and a system
architecture to prevent minors' exposure to numerous risks and
dangers while using Facebook, Twitter, and YouTube on a browser.
\end{abstract}



%
\IEEEpeerreviewmaketitle

\section{Introduction}\label{sec:intro}
The majority of teens ($85\%$) use more than one social media site according to a Pew Research Center \cite{anderson2018teens} survey ($N=743$). 
A 2018 poll ($N=1001$) \cite{Webpage:2019:Ofcom} found that the average 5 to 15 year-olds spend about 15 hours online every week. 
Additionally, $90\%$ of the 11 to 16 year-olds surveyed said that they have an online social network account. 
These numbers illustrate that the overwhelming majority of young people use OSNs, even if they are not old enough to legally register accounts for most mainstream OSNs, like Facebook, Instagram, Twitter, YouTube, and Snapchat.
Alarmingly, there are many risks adolescents are exposed to when using OSNs. 
Specifically, a 2019 study~\cite{Webpage:2019:NSPCC} of 21.6K primary school children and 18.1K secondary school children found that $16\%$ and $19\%$, accordingly, had seen content that encouraged people to hurt themselves. 
The same study reports that 11 to 18 year-olds reported seeing sexual content in the most popular OSNs. 
Last, reviews from over 2K young people aged 11 to 18, show that the $16\%$ witnessed violence and hatred, $16\%$ encountered sexual content, and the $18\%$ witnessed others being victims of cyberbullying. 
A different study conducted in 2018 found that $59\%$ of U.S. teens have
been victims of cyberbullying or harassment online.
Additionally, about a third ($32\%$) of teens report that someone has
spread false rumors about them on the Internet, while smaller
shares ($16\%$) have been the target of physical threats online.
Notably, the majority of the victims tend to be females. 
The study concludes that $59\%$ of the parents worry that their child might be getting bullied online, but most are confident they can teach their teen about acceptable online behavior~\cite{Webpage:2018:Pewresearch}.

Overall, the popularity of the Internet, and OSN usage in particular, is very high and with an increasing tendency among youngsters. 
Thus, the online risks for these sensitive age groups received increased awareness. 
To design an architecture for the protection of youngsters in OSNs, we list the most frequent dangers the young users might encounter. 
Existing literature~\cite{Webpage:2018:Pewresearch,Webpage:2017:Abc,Bookman:2016:Cyber} agrees to the following distinctive threats: i) cyberbullying; ii) cyberpredators; iii) sensitive information leakage; iv) manipulated content and pornography; and v) offensive images and messages.

\descr{Contributions.} In summary, this work makes the following contributions:
\begin{enumerate}
\item The design and implementation of a privacy-preserving CFAS that utilizes machine learning classifiers and other filters to protect minors when using OSNs.
\item CFAS makes efforts to keep the minors fully aware of what their custodians and what the Family Advice Suite can monitor,  filter, and analyze about their online activity.
\item CFAS employs fine-grained tools to spread awareness to the custodians and the minors about the various threats they face when using OSNs. 
It also utilizes the Guardian Avatar that interacts and advises the adolescents in a direct and user-friendly way.
\item The proposed architecture can accurately detect: (i) cyberbullying; (ii) sexual grooming; (iii) abusive users; (iv) bot accounts; (v) personal information exposure; (vi) sensitive content in pictures; (vii) hateful and racist memes; and (viii) disturbing videos.
\end{enumerate}

\descr{Paper Organization.} The rest of the paper is organized as follows.
First, we provide a detailed demonstration of the proposed architecture in Section~\ref{sec:architecture}, followed by our design principles in Section~\ref{sec:design}.
Then, we list and discuss how the classifiers hosted on the Intelligent Web-Proxy (IWP) work in Section~\ref{sec:implementation}.
We also provide an early evaluation of the system via a virtual environment, and physical experiments with beta testers (Section~\ref{sec:evaluation}), before discussing existing related work on parental control tools in Section~\ref{sec:relatedwork}.
Last, we conclude this work in Section~\ref{sec:conclusion}.

\section{Architectural Overview}\label{sec:architecture}

\begin{figure*}[t!]
\center
\includegraphics[width=0.95\textwidth]{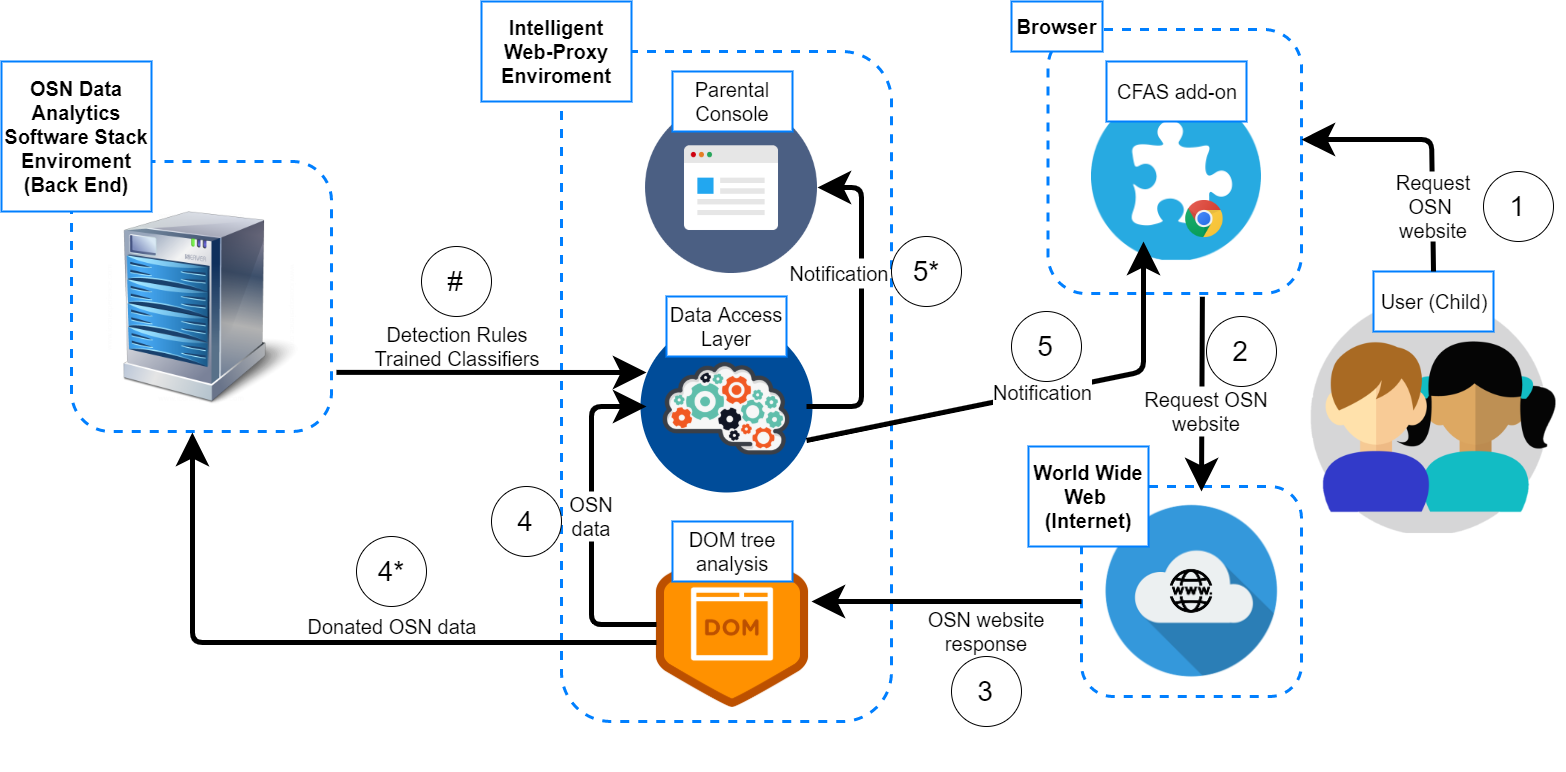}
\captionof{figure}{Cybersafety Family Advice Suite Architecture}
\label{fig:architecture}
\end{figure*}

In this section, we describe the main pillars of our architecture. 
This architecture comprises the following: 1) OSN Data Analytics  Software  Stack (Back-End); 2) Intelligent  Web-Proxy; and 3) browser add-on.  
For the tool to work efficiently,  all three components interact with each other, but none depends on the other to function.  
Figure~\ref{fig:architecture} depicts the proposed architecture of the  CFAS  framework, including its main components and the interfaces that interconnects them.
We describe the main purposes and functionalities of each component below.

\subsection{OSN Data Analytics Software Stack}
\begin{figure*}[t!]
\center
\includegraphics[width=0.95\textwidth]{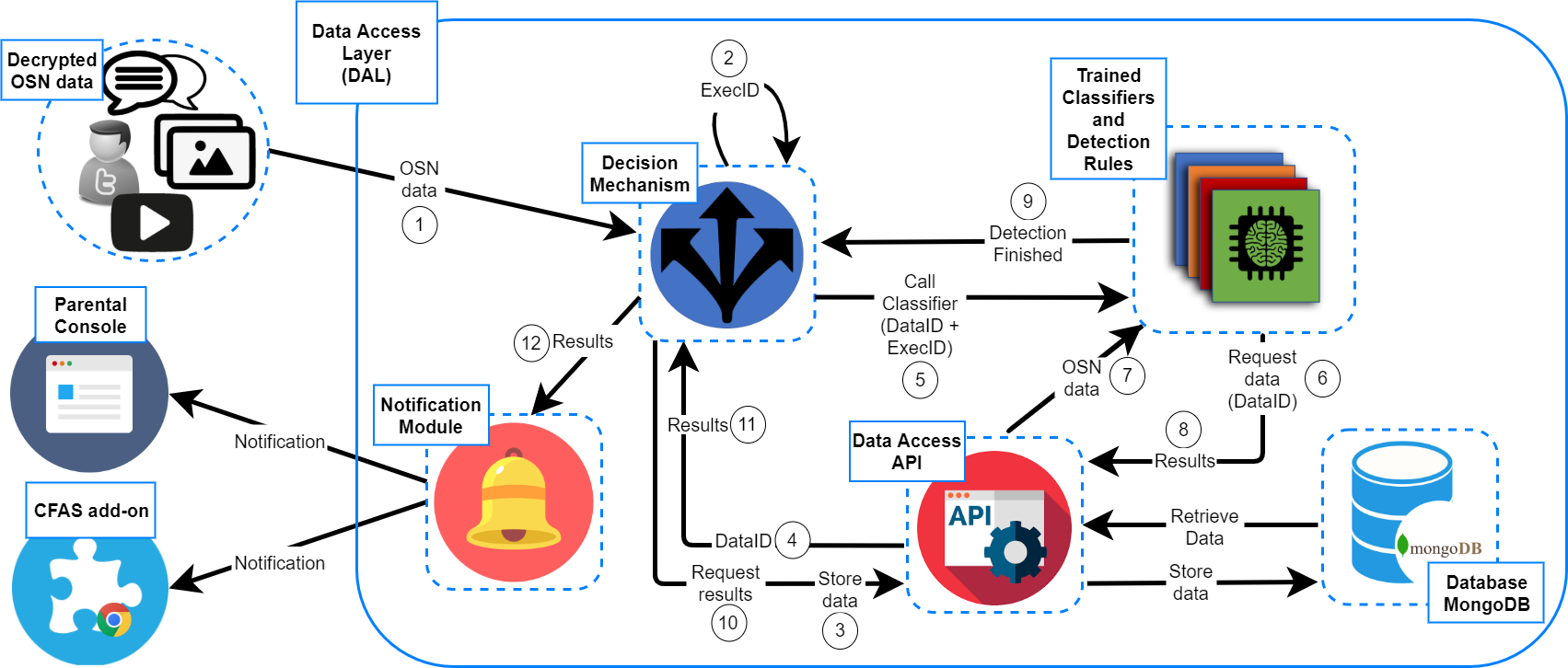}
\captionof{figure}{Data Access Layer (DAL) processes.  DAL is the main storage unit of the IWP and the Back-End of the CFAS infrastructure.}
\label{fig:dal}
\end{figure*}

The first component of the CFAS architecture is the OSN Data Analytics Software Stack, referred to as the \emph{Back-End} henceforth.
This is a single machine, which is responsible to train machine learning algorithms for the detection of threats in OSNs.
The trained classifiers and detection rules created on this machine are sent automatically to the registered Intelligent Web-Proxies (IWP) when available (see \# in Figure~\ref{fig:architecture}).
In addition, the Back-End stores anonymized OSN traffic data from the registered IWPs, \emph{only} if both the custodian and the minor give their explicit consent (4* in the figure).
These anonymized data are used to retrain the machine learning algorithms hosted in the Back-End to extract more accurate and intelligent classifiers, which are sent back to the IWPs to replace the existing classifiers, as shown in step \# in Figure~\ref{fig:architecture}. 

\subsection{Intelligent Web-Proxy}
The Intelligent Web-Proxy (IWP) is a small device that is connected to the router of the service provider in the house of the protected family.
We note that every different network needs its own IWP to be protected as a single IWP supports only one network.
The IWP consists of three modules that handle specific tasks, as described below.

\subsubsection{DOM Tree Analysis}
This part of the IWP captures all the incoming and outgoing traffic of the user (child).
Note that the word \emph{user} refers to the child protected by our architecture henceforth. 
First, the user requests a webpage using their browser (see 1 at Figure~\ref{fig:architecture}).
The response of this request is sent to the IWP: the DOM Tree Analysis module, specifically (step 3 in the figure).
After capturing the traffic, the DOM Tree Analysis module handles TLS connections and performs TLS termination to decrepit HTTPS websites (only Facebook and Twitter currently). 
Importantly, the IWP is tested to manage high network traffic load and extract the webpage content from the captured DOM tree. 
At the same time, the same data are sent to the Data Access Layer for analysis (see 4 in Figure~\ref{fig:architecture}).
We describe how the Data Access Layer (DAL) works below.

\subsubsection{Data Access Layer}
The Data Access Layer hosts all the trained classifiers and detection rules generated from the Back-End that are used to check all the received captured traffic.

Figure~\ref{fig:dal} demonstrates the functionality of the Data Access Layer, which is the main storage unit hosted in the IWP and the Back-End of the CFAS infrastructure. 
First, the data captured by the DOM Tree Analysis are sent to the \emph{Decision Mechanism} of DAL (step 1 in Figure~\ref{fig:dal}).
Every bit of information (Facebook chat, Facebook news-feed pictures, Facebook posts created by the user, Facebook pictures uploaded by the user, visited YouTube videos, and visited Twitter user profiles) is sent individually.
Upon reception of this data, the Decision mechanism creates a unique Execution ID (ExecID), see step 2 in the figure.
This unique string is used by the Decision mechanism to define the job number of the trained classifier, which is used to analyze the data.

Then, the Decision mechanism requests the Data Access API to store this data in the database: a MongoDB (step 3).
Once the data are stored, the Data Access API binds them with a unique number, which is used as a primary key to identify these data: DataID.
The DataID is sent back to the Decision mechanism (step 4), which is combined with the ExecID to call the suitably trained classifier to detect suspicious behavior (see step 5).
Once the trained classifier receives the ExecID and the DataID, it sends the DataID to the Data Access API to request the retrieval of data for analysis (step 6), which in return are sent back to the trained classifier (step 7).
Once the trained classifier finished the analysis of the data, it sends its results to the Data Access API, along with the ExecID and DataID to be stored in the database (step 8).
Then, the trained classifier sends the ExecID and DataID back to the Decision Mechanism to inform it that the analysis finished (step 9).

In response, the Decision Mechanism requests the results of the job from the Data Access API (step 10), and the Data Access API responds with the results of the analysis (step 11).
Last, based on the results of the trained classifier, and thresholds set in the Decision mechanism, the Decision mechanism is responsible to decide whether a notification needs to be sent to the user  via the CFAS browser add-on, and to the custodian of the user, via the Parental Console.
If this is the case, the Decision Mechanism triggers an event via the Notification Module (step 12).
Note that step 12 in Figure~\ref{fig:dal} is the same as step 5 and step 5* in Figure~\ref{fig:architecture}.

\subsubsection{Parental Console}
The last component hosted in the Intelligent Web-Proxy is the Parental Console.
The Parental Console is a fine-grained web-based platform that enables the custodian of the user to manage which data of the user (child) he/she and the IWP can see. 
Also, via the Parental Console, the custodian can choose what the IWP filters, protects, and blocks.
Additionally, custodians can set the level of the child's cybersafety. 
To set these options in operation, the child receives notifications on their browser add-on through the Notification Module, informing them that their custodian has made some changes in the options.

We highlight that for these options to operate, the child needs to approve them via their browser add-on.
This way, we ensure that the child gave their consent about what the IWP captures, analyzes, filters, and blocks. 
At the same time, this functionality ensures that the child knows exactly what notifications their custodian will be receiving about the online activity of the child, and what OSN traffic activity the custodian can see.
We note that our proposed architecture promotes a conversation and close communication between the custodian and the child.
This way, the family protected by CFAS can agree on what online activity of the child the custodians need to monitor, and what are the main risks and threats involved in using OSNs.
Moreover, this architecture promotes OSN threat awareness, hence enforcing a culture of safe OSN usage.
To achieve this, we introduce specific Parental and Back-End visibility options and Cybersafety options.

\par 1) Parental Visibility Options: These options define what the custodian of the user can see, while enabling various levels of monitoring for the custodians, always with the explicit consent of the user. 
We define three Visibility Levels: 
\begin{itemize}
    \item{Level 1:} This is the lowest level of parental visibility, meaning that the custodian cannot see any data regarding the OSN traffic of the user.
    We note that the custodian still receives notifications regarding the threats detected by the trained classifiers hosted in the IWP, without mentioning the name of the perpetrator or revealing any OSN data.
    For the sake of the following examples, we assume that the protected child's name is \emph{John}: ``John might be a victim of cyberbullying.''
    \item{Level 2:} This level of visibility allows the custodian to select some of the following OSN activity of the child to be visible to them: suspicious Twitter usernames the child visited, disturbing YouTube videos the child watched, Facebook wall, photos, and friends of the child. 
    Once the user gives their consent via their browser add-on for this data to be visible to the custodian, the visibility option is operational.
    A notification example: ``John might be a victim of cyberbullying by Eve'', where John is the protected child, and Eve is the perpetrator.
    \item{Level 3:} This is the default and highest level of parental visibility. 
    When this option is selected, it adds all the options from Level 2, along with data regarding the user's Facebook chat. 
    So at this level, the custodian of the child can see all the incoming and outgoing traffic of the child's Facebook wall, photos, notifications, friends, and chat, \emph{only} in case of an incident. 
    A notification example: ``John might be a victim of cyberbullying by Eve. Click here to see the suspicious chat''.
    This way, the custodian can see \emph{portions} of the chat between the user and the perpetrator that show signs of cyberbullying.
\end{itemize}
We note that these options expire once every six months, so the custodian and the child can reset them as they wish. 
All the above levels of visibility can be set up after a mutual agreement between the custodian and the user while keeping the user fully aware of what their custodian can see.

\par 2) Back-End Visibility options: Through the Back-End Visibility options, the Cybersafety Family Advice Suite offers options regarding which OSN traffic data is sent to the Back-End.
OSN data sent to the Back-End are used to retrain the machine learning algorithms and detection rules hosted there to make them more accurate in future predictions.
The custodian can choose among the child's Facebook wall, photos, notifications, friends, and chat.
We note that the user needs to give their consent for the data to be sent to the Back-End.
We define the following Back-End Visibility Levels:
\begin{itemize}
    \item{Level 1:} This is the lowest level of Back-End visibility. 
    If this option is set, no data is sent to the Back-End. 
    \item{Level 2:} In this level, the custodian allows the IWP to send data to the Back-End regarding the child's Facebook wall, friend's Facebook wall, and the child's Facebook friends profiles.
    The custodian may select one or all of the above. 
    Also, these data may be sent anonymized or not.
    \item{Level 3:} This is the highest level of Back-End visibility.
    When this option is set, it allows the IWP to send all the data from level 2, in addition to the child's Facebook chats. 
    Once again, these data may be sent anonymized or not, and always with the consent of both the custodian and the child. 
\end{itemize}

\par 3) Cybersafety Options: Last, the Parental Console allows the custodian to choose the child's level of Cybersafety. 
These options define how aggressive the IWP can be, regarding the protection of the user: what the IWP can filter, protect, block, replace, encrypt, or watermark.
This options can be configured at two different levels: 
\begin{itemize}
    \item{Level 1:} This is the lowest level of cybersafety.
    If set, the IWP only pushes notifications to the user explaining that certain suspicious or malicious activity is detected.
    This means that the IWP still detects suspicious activity, but it does not hide, protect, encrypt, blocks, or watermarks any content.  
    Via the Parental Console, the custodian can choose the notifications they wish for the child to receive for each detection mechanism.
    The detection mechanisms include: a) cyber grooming; b) hate or inappropriate speech (cyberbullying); c) distressed behavior (when the child is suicidal, scared, depressed); d) fake activity (fake OSN profiles); e) personal information exposure (when the child is about to publish personal information); f) hateful memes; g) inappropriate YouTube videos; and h) sensitive content in pictures (when the child is about to share a benign picture that includes nudity without protection, like a picture in a swimsuit).
    \item{Level 2:} At this level, the custodian may choose any of the above IWP detection mechanisms to take action and filter, replace, protect, encrypt, or block content before it reaches the browser of the protected child.
    The detection mechanisms remain the same as level 1, but the custodian needs to select at least one to be operational for this level to hold.
\end{itemize}

Overall, the IWP is responsible for capturing the incoming and outgoing traffic of Facebook, Twitter, and YouTube of the user and send it to the locally hosted trained classifiers to detect malicious activity. 
In case the suspicious activity is detected by one or more trained classifiers, the IWP pushes a notification to the browser add-on of the user to inform them about the imminent threat detected.
At the same time, the suspicious malicious content is blocked or filtered by the browser add-on to protect the minor, given that the Cybersafety Option Level 2 is set by the custodian and the user.
The IWP hosts trained classifiers and detection rules to perform the following actions:
\begin{enumerate}
    \item detect nudity in images included in the captured traffic; 
    \item encrypt sensitive images with steganography; 
    \item detect and warn the minor in case they are about to share personal information; 
    \item detect cyberbullying in Facebook conversations; 
    \item detect sexual grooming in Facebook conversations; 
    \item detect hateful and racist memes in Facebook feed; 
    \item detect bot, aggressive, bully, and spam Twitter users; 
    \item detect inappropriate videos for children on YouTube; 
    \item provide sentiment analysis of the chat of the minor; 
    \item generate informative notifications to the minor; 
    \item push notifications to the custodian about an incidence (e.g., sexual grooming); 
    \item push notifications to the child via the browser add-on; 
    \item submit data to the Back-End through a secure tunnel; and 
    \item block adult, or any other site, defined by the custodian.
\end{enumerate}

\subsection{Browser add-on}
The last component of our architecture is the browser add-on (CFAS add-on in Figure~\ref{fig:architecture}).
The browser add-on is the gateway between the IWP and the user, responsible to inform the user about the threats detected from the IWP, and the Visibility and Cybersafety options set by their custodian.

Importantly, our browser add-on operates as a Guardian Avatar that the child may interact with to ask for advice.
Our avatar operates as the \emph{guardian angel} of the user while using different OSN platforms (Facebook, Twitter, and YouTube only, currently).
By following the Guardian Avatar approach as a gamification  feature \cite{Deterding2011}, CFAS aims to encourage the users to use it and interact with it because of its extended usability and improved user experience functionalities. 

In addition, the user can select their favorite avatar icon from a list of icons. 
The Guardian Avatar ``follows'' the user in their online-activities as a virtual friend. 
When the IWP detects any malicious behavior or incidents, the notifications (warnings, advice, etc.) appear as chat bubbles of the avatar, in a friendly and encouraging text.
An example of the avatar notifying the minor about a detected incident is depicted in Figure~\ref{fig:guardian_avatar_example}.
With the addition of the avatar, it is expected that the CFAS warnings and advice will be less disturbing for children (especially for the adolescents) and will make users more willing to use it.
\begin{center}
\includegraphics[width=0.95 \columnwidth]{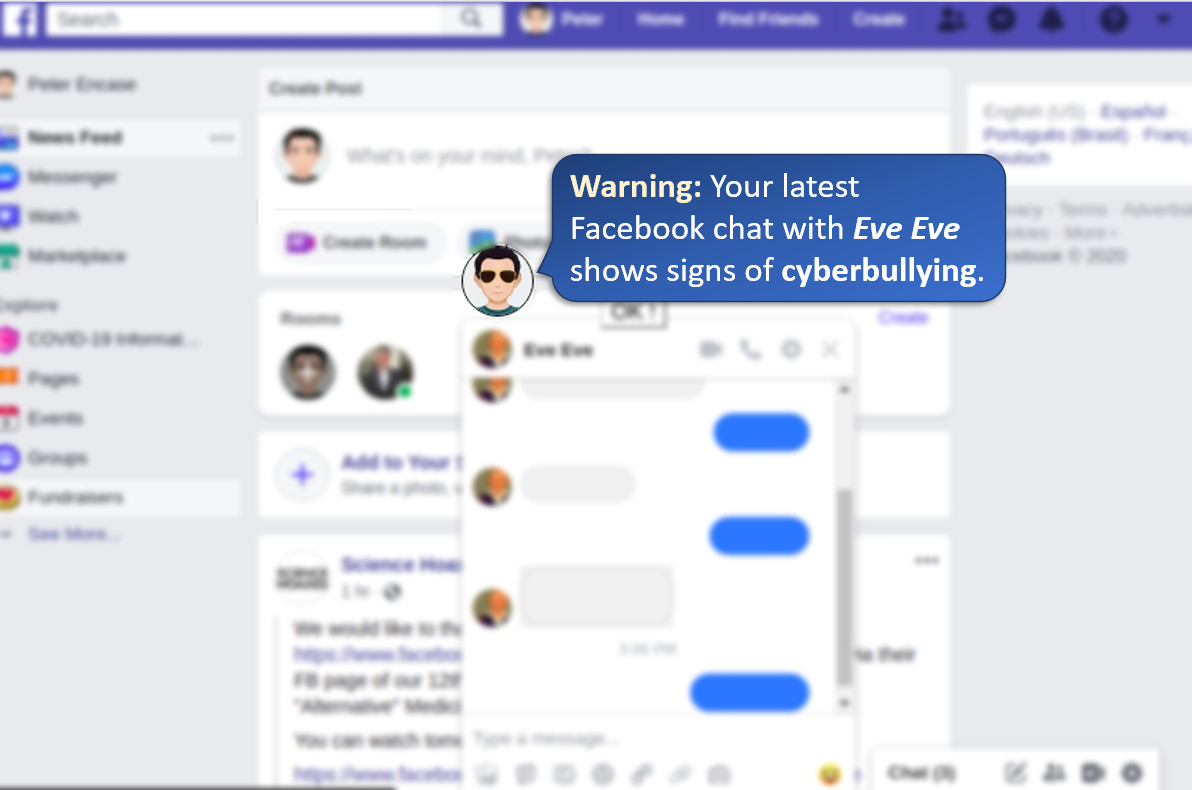}
\captionof{figure}{Guardian Avatar notifies the minor of any detected incidents}
\label{fig:guardian_avatar_example}
\end{center}

\noindent{The browser add-on can:}
\begin{enumerate}
    \item notify the user about the activity detected by the IWP;
    \item notify the user about what their custodian can see based on the preferences (Parental Visibility options) applied;
    \item notify the user about what data is sent to the Back-End to aid the machine learning classifiers to become more accurate (Back-End Visibility options);
    \item let the user change the options about what OSN traffic activity their custodian can see;
    \item let the user change the options about what data is sent to the Back-End;
    \item let the user flag content/text as cyberbullying activity, sexual cyber grooming activity, aggressive behavior activity, fake identity activity, and false information activity in case the IWP failed to detect so;
    \item let the user flag sensitive or nudity content in case the IWP failed to detect so; and
    \item let the user flag content/text as an incorrect sensitive content, cyberbullying, sexual grooming, aggressive behavior, fake identity, and false information activity in case the IWP detected so. 
\end{enumerate}

Overall, we propose a fully privacy-preserving architecture for the protection of minors when they use OSNs, both towards their custodians and towards the system itself.
First, the minor is empowered to choose the online activity and warnings that their custodian receives in case a threat is detected by the IWP.
This can be done via the Parental Visibility Options.
Second, the user can choose which online activity the IWP filters, captures, and protects via the Cybersafety Options.
Also, the IWP, the device that is responsible for capturing and analyzing the online activity of the minor to detect online threats, is connected and physically exists within the network of the user.
Thus, the online activity of the minor is captured and analyzed locally and is isolated within the network of the user.
In addition, the IWP never makes any data visible to the rest of the system (Back-End or other IWPs) without the explicit authorization and consent of both the user and their custodian via the Back-End Visibility Options.

\section{Design}\label{sec:design}
We now detail the design of the proposed architecture. 
Instead of simple rule-based filters, our architecture utilizes advanced machine learning algorithms. The downside of having rule-based filters is that they are blunt. 
There are situations where there is a particular piece of content that technically does not violate the specified policies, but when this content is analyzed with advanced machine learning techniques, it might turn out to be hate speech, sarcasm, sexual grooming, etc. 
Such techniques allow us to detect bullies or predators that are close to the line. 
To sum up, the aim is to have these granular standards so that our design can control for bias.
Our design approach is based on the following design principles:

\par 1) We place all functionalities (filters, text replacement, notifications, data submission to the Back-End, etc.) in the IWP instead of the browser add-on when it can be correctly and efficiently implemented. 
This way, we prevent a minor from modifying or disabling the system's functionality through the browser add-on.
For example, in case a minor accidentally or willingly disables the browser add-on, the IWP does not get affected, and all the processes and functionalities can continue their operation normally.
We assume that the device of the minor is still configured to route social network services through the IWP and that the child does not have the permission, knowledge, or access to alter the configuration of the IWP or their personal device. 
Also, the IWP can notify the custodian through the Parental Console that the browser add-on of the minor is not responding anymore. 

This architecture aims to provide the ability to seamlessly support multiple types of clients (desktop browsers, mobile apps, etc.) with a minimal client or client platform configurations or modifications.
Moreover, the browser add-on does not support complex functionalities other than javascript and HTML scripts.
For example, functionalities, like text replacement, picture encryption, filtering, etc., are too complex to be implemented and run on a browser add-on. 
 
In case the IWP is down, the browser add-on calls REST API requests from the Back-End, and the Back-End DAL is employed to identify suspicious content.
This means that the OSN traffic activity of the user is sent outside of the network, to the Back-End, for analysis.
Whether a suspicious activity is detected by the Back-End or not, all the user OSN traffic data is automatically deleted from the Back-End.
Having some functionalities on the IWP prevents it from calling REST API requests from the Back-End every time it needs to analyze OSN traffic activity. 
In addition, placing some functionalities on the IWP, solves the potential problem of the whole system being down in case of Back-End unavailability, thus solving the problem of single-point failure. 
Examples: i) The IWP can push notification to the browser add-on without the need of the Back-End. ii) Before any content reaches the minor's device, the IWP can replace cyberbullying content without calling REST API requests from the Back-End, using the functionality installed on it already.

\par 2) Rules and trained classifiers are generated in the Back-End.
Trained classifiers are placed in the IWP only if they can run efficiently. 
The Back-End collects data from all the IWPs to generate detection rules and trained classifiers. 
Data collected from the IWPs are used to generate cyberbullying, sexual cyber grooming, distressed behavior, aggressive behavior, fake identity, and false information detection rules.

\par 3) Warning, flagging, and feedback functionality is placed on the browser
add-on.
The Guardian Avatar displays notifications in dialogue boxes after the IWP detects suspicious behavior and pushes a notification to the browser add-on.
The user can flag content as cyberbullying activity, sexual cyber grooming activity, aggressive behavior activity, fake identity, false information, and sensitive picture through the browser add-on in case the IWP failed to detect so. 
The user can also give feedback based on the activity detected by the IWP. 
For example, in case the IWP detects cyberbullying, it pushes a notification to the browser add-on. 
The Guardian Avatar shows the notification/warning to the user explaining that cyberbullying was detected (Figure~\ref{fig:guardian_avatar_example}). 
Then, the user can provide feedback on whether this detection is accurate or not. 

\par 4) The minor can check the content their custodian, the IWP, and the Back-End can see.
The custodian can set up the Visibility settings in a fine-grained way and always with the consent of the minor.
This way, we enable various levels of monitoring for parents and the Back-End with the child's consent, while keeping the child fully aware of what their custodians and the Back-End
can see, e.g., chat messages.

Overall, we propose a system that eases the tension of ensuring the safety of minors while respecting their privacy with respect to what their custodians and third parties can see. 
By automating the detection of malicious communication, we enable custodians to be
continuously aware of their child's safety. 
This is achieved without the parent having to go through the minor's online communication manually, thus, without having to invade the minor's privacy. 
Our approach aims to warn the custodians about the suspicious online activity that was detected, without violating the privacy of the minor.
For example, if the minor has a Facebook online conversation with sexual content with somebody, the custodian of the minor will receive a warning that such a conversation is taking place, once the IWP captures it. 
Still, the parent won't be able to see the actual content because that would violate the teenager's privacy. 
Instead, the parent can only see the actual conversation through their Parental Console once the explicit consent of the child has been granted. 
To sum up, our design principles intend to encourage custodians to have a conversation with the minor; thus, bringing families closer and spreading awareness about the numerous threats that exist in contemporary OSNs.

\section{Implementation}\label{sec:implementation}
We implement all the architecture components, and integration's that we describe in Sections~\ref{sec:architecture} and~\ref{sec:design}.
In this section, we provide the details of the prototype implementation.
Note that we employ classifiers created in previous work for the detection of threats in OSNs.
We note that these classifiers are generated on the Back-End and hosted on the IWP.
In case the classifiers detect suspicious activity, the IWP pushes notifications to the browser add-on of the user, and the Parental Console.

\subsection{Detection of Abusive Users on Twitter}
When the minor visits a Twitter user account, the IWP captures the username of the visited user, and it calls the Twitter API to collect the last 20 tweets (including retweets) of that user~\cite{Webpage:2020:TwitterAPI}.
This information is then sent to a classifier developed by Chatzakou et al.~\cite{Bookman:2017:Mean} for analysis.
The developed classifier is trained with Twitter annotated data~\cite{twitter1}\cite{twitter2} and analyzes the last 20 tweets of the visited Twitter user to detect whether it is an aggressive, bully, spam, or normal account.

\subsection{Fake and Bot user detection on Twitter}
When the minor visits an account on Twitter, the IWP captures the username of the Twitter account and sends it for analysis via a REST API call developed by~\cite{Webpage:2019:astroscreen} and Echeverria et al.~\cite{Bookman:2018:LOBO}.
This API returns True if the Twitter user account is a bot, and False otherwise.
In case of the former, the IWP pushes a notification to the browser add-on of the minor, and to the Parental Console of the custodian (based on the Parental Visibility options).

\subsection{Detection of Hateful and Racist memes on Facebook}
The IWP captures the Facebook incoming and outgoing traffic of the minor and performs TLS termination of the DOM tree. 
All the images that are extracted from the DOM tree are sent to the classifier developed by Zannettou et al. \cite{Bookman:2018:Origins} to be labeled as a hateful meme or not. 
This classifier is trained using images from Twitter, Reddit, 4chan’s Politically Incorrect board~\cite{papasavva2020raiders}, and Gab~\cite{memes}.
In case the detection is positive, the picture will be automatically replaced by the IWP with a static image to inform the minor.

Similarly, when the minor uploads an image on Facebook, the picture is analyzed by the aforementioned classifier to detect whether that image is hateful or racist.
If so, then the IWP pushes a notification to the guardian avatar to advise the minor that the image they try to upload contains hateful content, and they shouldn't upload it.

\subsection{Sexual Predator Detection on Facebook}
When the minor is chatting with a friend on Facebook, the conversation is captured by the IWP and is sent to the classifier developed by Partaourides et al.~\cite{partaourides2020self} for analysis.
A previous version of this classifier was trained with data from Perverted Justice website~\cite{Webpage:2019:PervertedJustice} to recognize patterns similar to the ones from convicted sexual predators.
Upon positive detection, the IWP pushes a notification to the browser add-on of the minor, notifying them that signs of sexual predator have been detected.
The custodian can see only portions of the chat between the minor and the predator via the Parental Console, only if the minor consents so via the Parental Visibility options explained in Section~\ref{sec:architecture}.
We note that the custodian can only see portions of the chat that the classifier detects as a sexual grooming pattern.

\subsection{Cyberbullying Detection on Facebook}
Similar to the Sexual Predator detection, when the minor is chatting with a friend on Facebook, the conversation is captured by the IWP and is sent to the classifier developed by Partaourides et al.~\cite{partaourides2020self} for analysis. 
This classifier returns percentages of how angry, frustrated, and sad the minor is during the Facebook chat conversation, using sentiment analysis.
If any of these three feelings exceed $65\%$, the IWP pushes a notification to the browser add-on of the child to warn them that the Facebook chat they are having seems to be toxic for them. 
Similar to the sexual predator detection above, the custodian is only able to see portions of the suspicious chat, only if the minor gave their consent beforehand. 

\subsection{Personal Information Leakage Detection on Facebook}
When the user tries to make a post on Facebook, the IWP captures the text written by the user and analyzes it to detect dates, times, phone numbers with or without extensions, links, emails, IP and IPv6 addresses, prices, credit card numbers, street addresses, and zip codes.
We implement this detection technique using existing Python libraries \cite{Webpage:2019:Regex}.
In case any of the above personal information is detected, the IWP pushes a warning to the minor to remove the sensitive information from their post.
In case the minor dismiss these warnings, a notification is sent to the Parental Console of the custodian (in accordance with the Parental Visibility options).

\subsection{Watermarking and Steganography}
For the purposes of this detection mechanism, we consider any image that includes nudity (topless images of boys, or swimsuit images) as sensitive content images.
When the minor tries to send a sensitive image to a friend over Facebook chat, the image first passes in the IWP for analysis. 
We followed similar techniques to Ghazali et al.~\cite{osman2012enhanced} and Kolkur et al.~\cite{kolkur2017human} to develop our skin and nudity detection techniques.
In case the image contains sensitive content, the IWP watermarks it~\cite{Webpage:2005:watermark}.
Then, the IWP hides the original image in another static image using steganography.
This way, only the person that the picture was sent to is allowed to see the hidden original image.
We note that for this to work, the receiver needs to be part of the Cybersafety Family Advice Suite network as decryption keys hosted on the Back-End are requested from the IWP to decrypt the image.
Similarly, if the minor tries to post an image that contains sensitive content on their Facebook wall, the IWP watermarks and performs steganography techniques to the image before posting it on Facebook.
The minor, using the browser add-on, can set who is able to see (decrypt) this picture (family members, friends, classmates, etc.).
For this scenario, we assume that the minor allows the image to be visible to family members only, and that their family members are registered CFAS members and have their own IWP set up at home.
When a family member of the minor scrolls Facebook, their IWP captures that image and communicates with the CFAS Back-End to check if they have permission to see this image.
If this is the case, then the IWP decrypts the image automatically.
In case the image does not contain sensitive content, the IWP only applies watermarking on it before posting it.
The receivers that are not part of the CFAS network can only see the static encrypted image.

\subsection{Disturbing videos on YouTube}
Our architecture also detects disturbing YouTube videos for young children, using the developed classifier by Papadamou et al.~\cite{Bookman:2020:Youtube}. 
This classifier was trained using YouTube videos~\cite{disturbedyoutube} and can discern inappropriate content with $84.3\%$ accuracy. 
When a minor visits a YouTube video, the IWP captures the YouTube link, which includes the YouTube video ID, and it calls the YouTube API to collect the video features~\cite{Webpage:2020:youtubeAPI}.
These features include the video upload date, likes, tag, title, thumbnail, etc. 
The IWP then sends these video features to the developed classifier for analysis.
In case the classifier returns positive detection (inappropriate), then it warns the minor that the video they are watching is not suitable for them via the browser add-on.

\section{Evaluation}\label{sec:evaluation}
In this section, we evaluate the performance of the prototype implementation of the Cybersafety Family Advice Suite.

\subsection{Performance Evaluation}
To test the performance in regard to the number of concurrent users, we set a small home cluster using a laptop with 4GB Ram, a quad-core Intel Core i5 processor that is running Ubuntu 18.04 64bit and Google Chrome Version 80.0.3987.162 (64 Bit), which is used as the minor's laptop that hosts the browser add-on.
In addition, we set up two virtual machines with 2GB RAM each, and one tablet of 3GB RAM: 4 users in total.
The IWP is a virtual machine hosted on the Google cloud, configured with 4GB RAM, a dual-core Intel Xeon CPU, running Centos 7 (64 Bit), and it is using the mitmproxy~\cite{Webpage:2020:mitmproxy}: the HTTPS proxy. 
Also, the IWP hosts a MongoDB for Data storage and Python3 for the API Calls. 
We run the experiments with a downlink of $\sim$20 Mbps and an uplink of $\sim$5 Mbps.

\begin{figure}[t!]
\center
\includegraphics[width=1\linewidth]{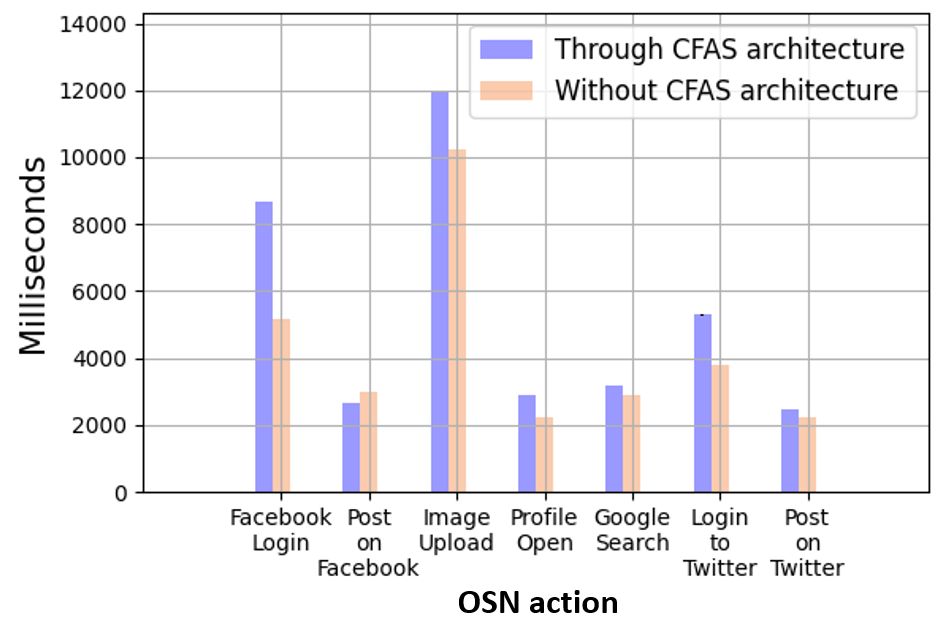}
\captionof{figure}{OSN actions with CFAS \& without CFAS}
\vspace{-4mm}
\label{fig:osn_actions_with_without_cfas}
\end{figure}

Figure~\ref{fig:osn_actions_with_without_cfas} depicts the time in milliseconds needed for OSN actions to be executed with and without CFAS.
Each machine executes the OSN actions using a JavaScript automated method in a serial manner.
Then, we calculate the average time that each machine needed to finish each action using the start time and end time of each action.
We observe that with CFAS, there are reasonable delays regarding the execution of some actions (e.g., Facebook Login, Image Upload, Twitter Login). 
This delay is acceptable since extra processing is needed to load and execute the CFAS tools. 
Other actions' delay is negligent ($\sim$1 second).

\subsection{User Experience}
In this section, we present the results of a user experience evaluation questionnaire given to minors and custodians after interacting with CFAS. 
The participation of minors required their custodians' consent. 
The sample consists of 30 minors and 12 custodians that had no knowledge or experience of the CFAS tools. 
The questionnaires were GDPR-compliant and anonymous. 
The study has received data protection approvals by the Ethics Committee of the Cyprus University of Technology, and by the Office of the Commissioner for Personal Data Protection of the Republic of Cyprus.

To evaluate our tools, the minors had to answer a variety of questions regarding their usability, accessibility, and performance. 
The minors were between 12 to 16 years old and reported using the Internet daily for entertainment and education purposes.
The percentages of minors in our sample that have a registered Facebook, Instagram, and YouTube account are $53.3\%$, $33.3\%$, and $13.3\%$, respectively.

\begin{figure}[t!]
\center
\includegraphics[width=0.45\textwidth]{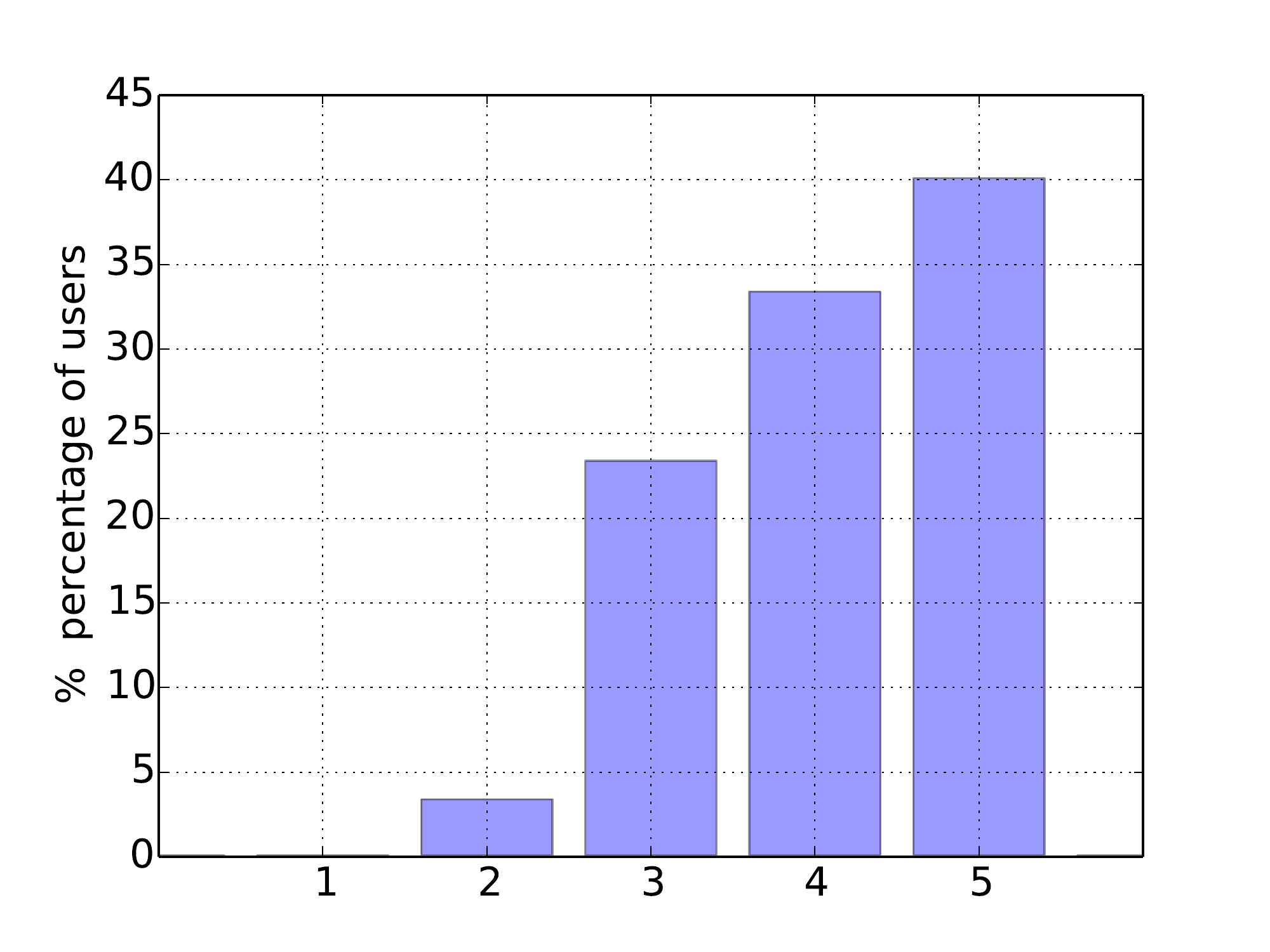}
\vspace{-1em}
\caption{(Minors) Would you allow CFAS to send notifications to your custodian regarding suspicious detection? (1: Totally Disagree, 5: Totally Agree)}
\vspace{-1.5em}
\label{fig:minor_consent}
\end{figure}

\begin{figure}[t!]
\center
\includegraphics[width=0.45\textwidth]{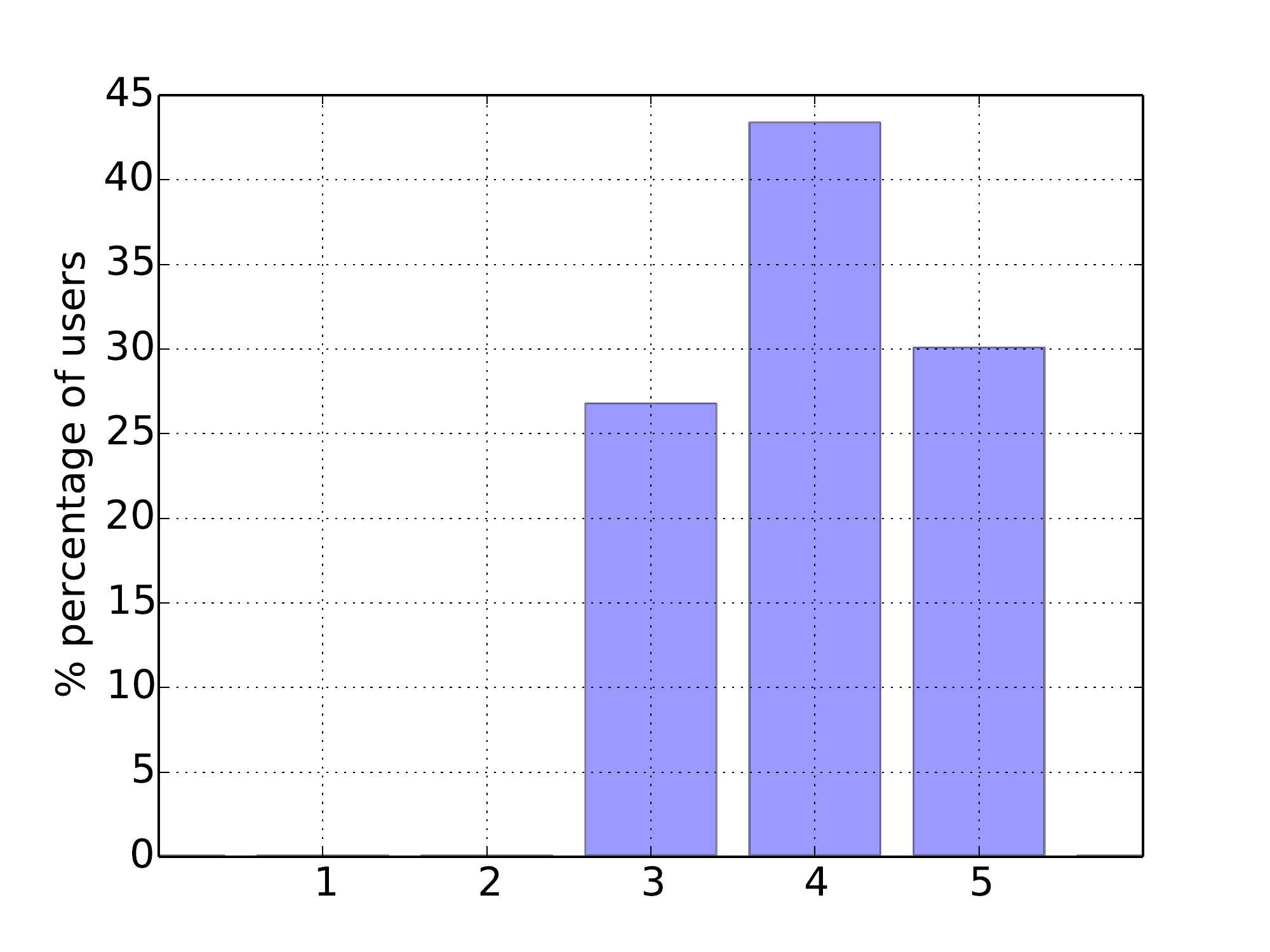}
\vspace{-1em}
\caption{(Minors) Do you believe CFAS would improve your safety when using OSNs? (1: Totally Disagree, 5: Totally Agree)}
\vspace{-2em}
\label{fig:safety_improvement}
\end{figure}

\begin{figure}[t!]
\center
\includegraphics[width=0.45\textwidth]{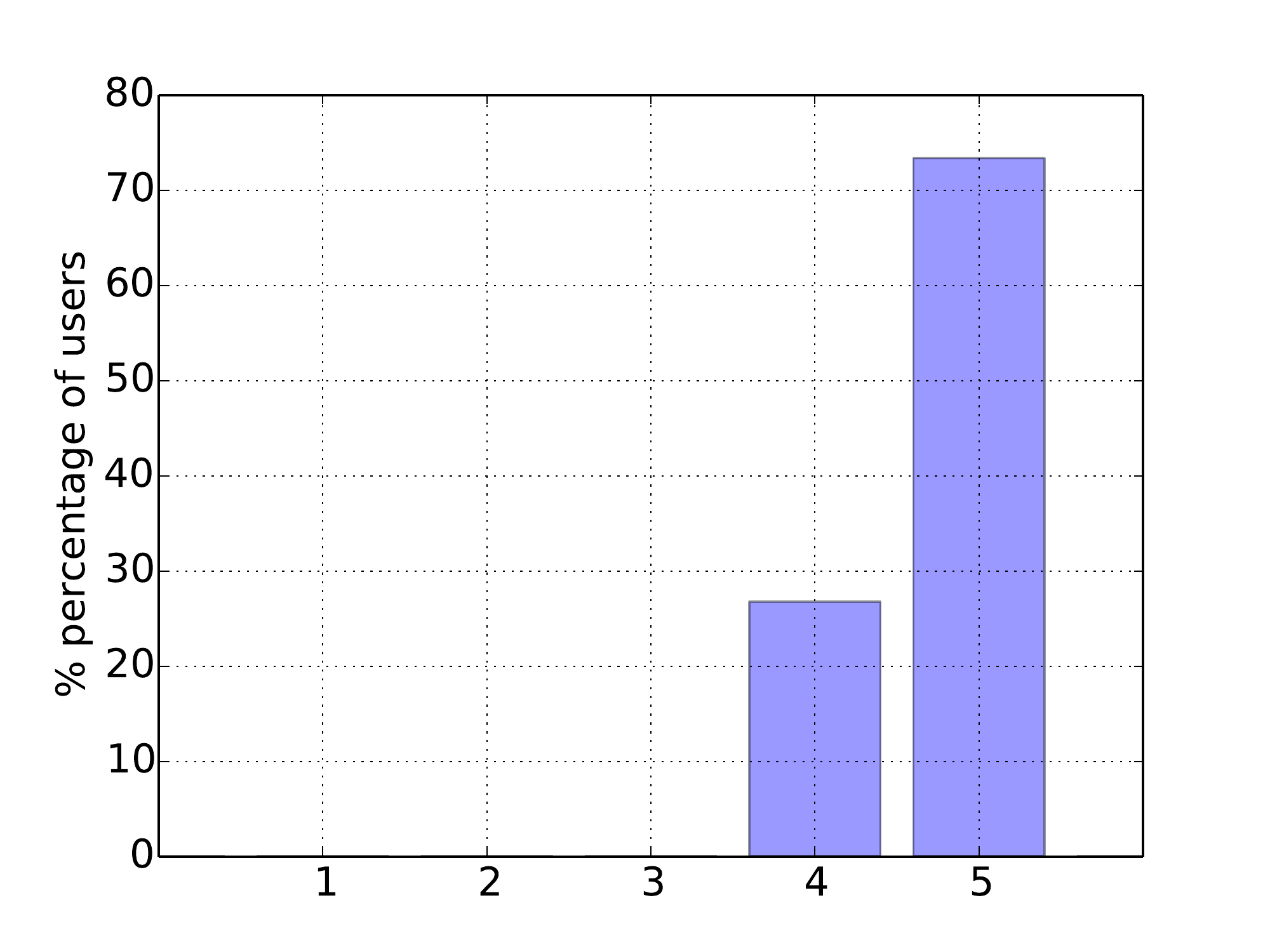}
\vspace{-1em}
\caption{(Minors) Are you satisfied with CFAS capabilities? (1: Totally Disagree, 5: Totally Agree)}
\vspace{-2em}
\label{fig:capabilities}
\end{figure}

\begin{figure}[t!]
\center
\includegraphics[width=0.45\textwidth]{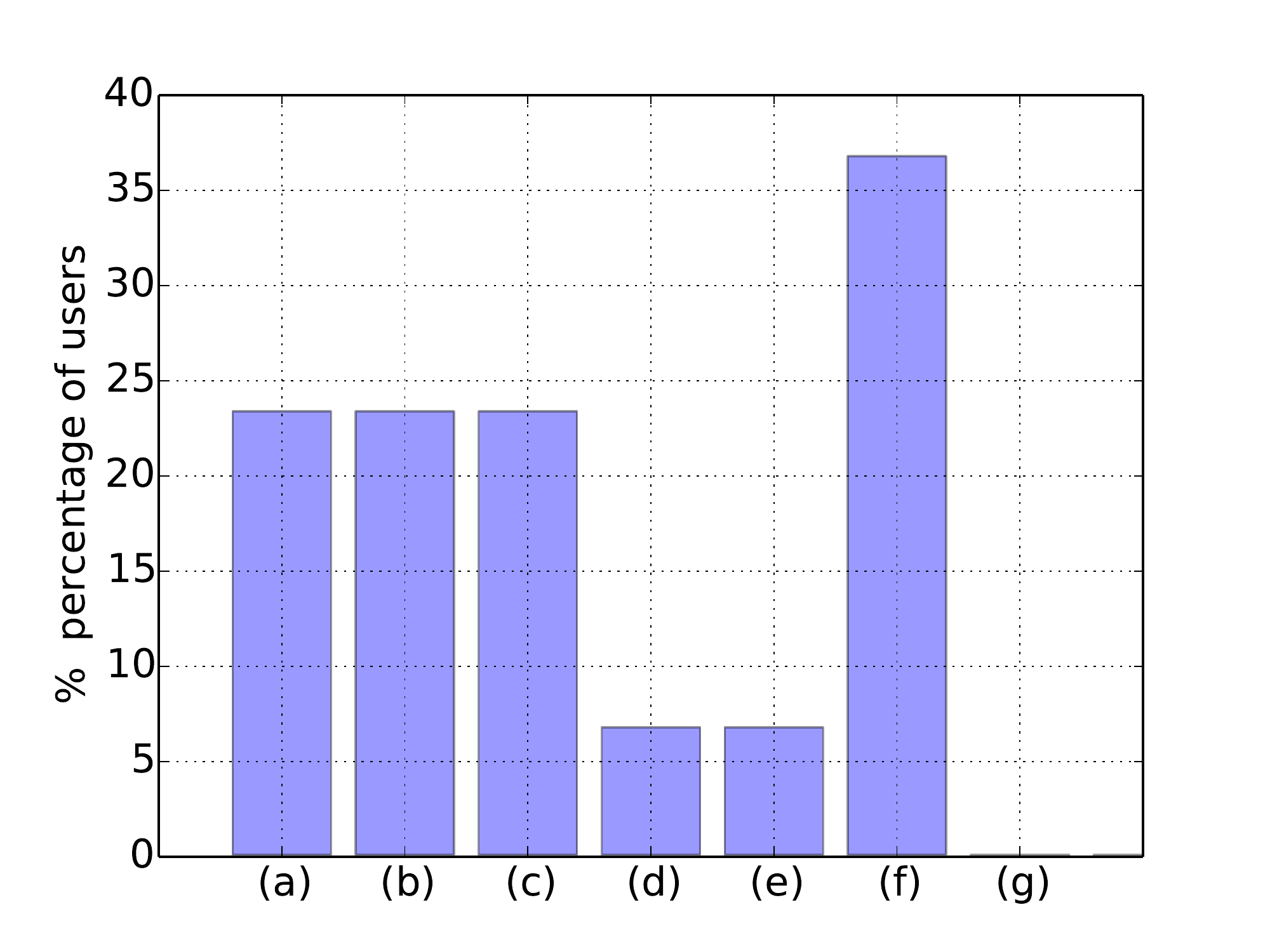}
\vspace{-1em}
\caption{(Minors) Have you ever experienced the following online-threats? Select all that apply to you: (a) I prefer not to say; (b) None; (c) Personal data misused; (d) Personal photo misused; (e) Cyberbullying; (f) Inappropriate speech and racism; and (g) Sexual grooming}
\vspace{-2em}
\label{fig:osn_threads}
\end{figure}

We report some of the results we obtained from the questionnaires given to minors and their custodians after they used the CFAS tools.
When minors asked whether they would allow CFAS to send notifications to their custodians, the majority reported high, and complete agreement (Figure~\ref{fig:minor_consent}).
In addition, the majority of minors believe that these tools could improve their safety when using OSNs, as depicted in Figure~\ref{fig:safety_improvement}.
Importantly, all of the minors report being very happy with the capabilities of CFAS (Figure~\ref{fig:capabilities}).
Alarmingly, Figure~\ref{fig:osn_threads} depicts that many minors had their personal data ($24\%$) and photos ($7\%$) misused, being a victim of cyberbullying ($7\%$), and witnessing inappropriate speech and racism ($37\%$) on social networks.
Note that the minors could select any that applied to them for this question.

\begin{figure}[t!]
\center
\includegraphics[width=0.45\textwidth]{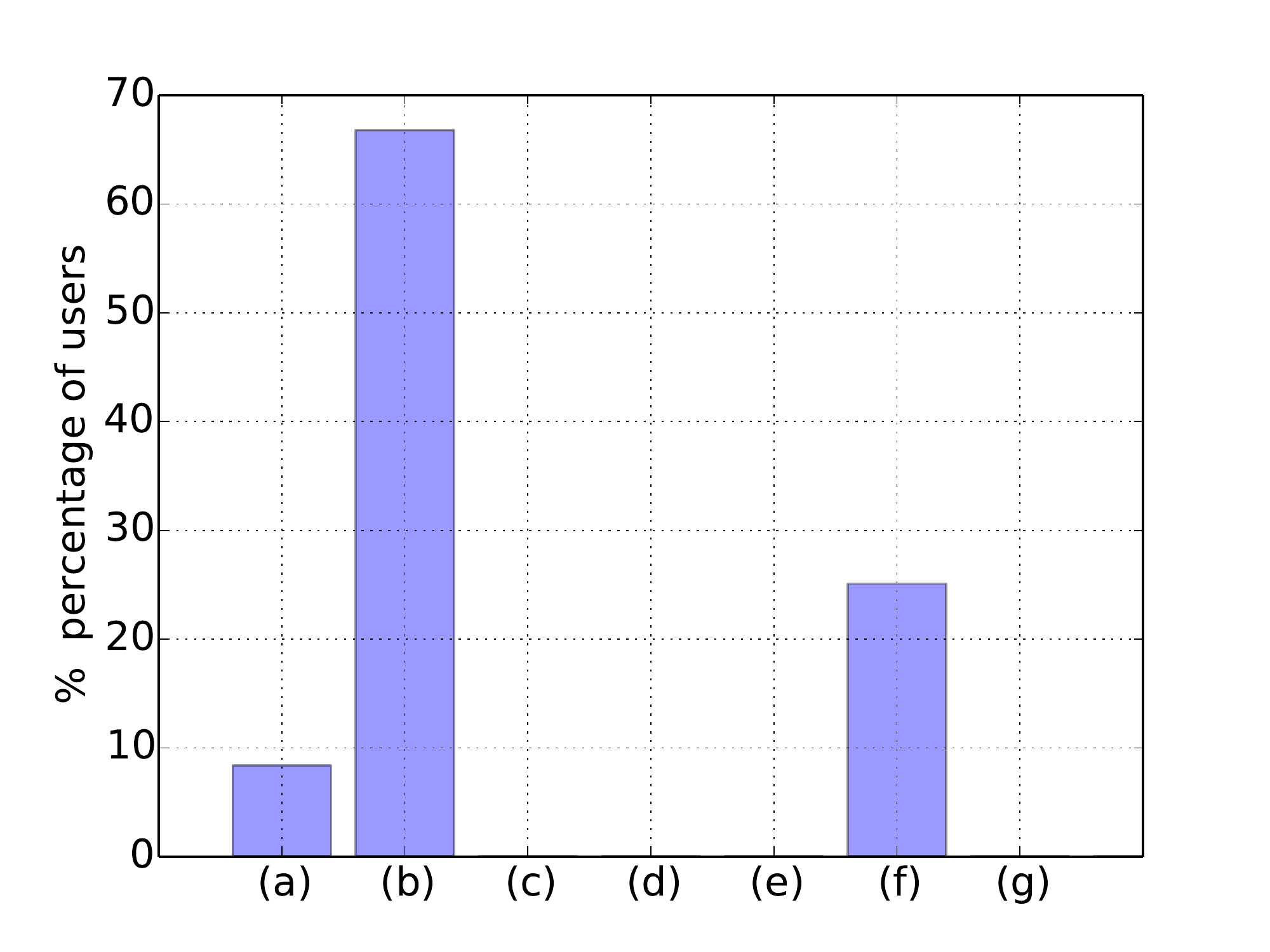}
\vspace{-1em}
\caption{(Custodians) Has your child ever reported to you being a victim of the following? (a) I prefer not to say; (b) None; (c) Personal data misused; (d) Personal photo misused; (e) Cyberbullying; (f) Inappropriate speech and racism; and (g) Sexual grooming}
\vspace{-2em}
\label{fig:custodian_osn_threats}
\end{figure}

\begin{figure}[t!]
\center
\includegraphics[width=0.45\textwidth]{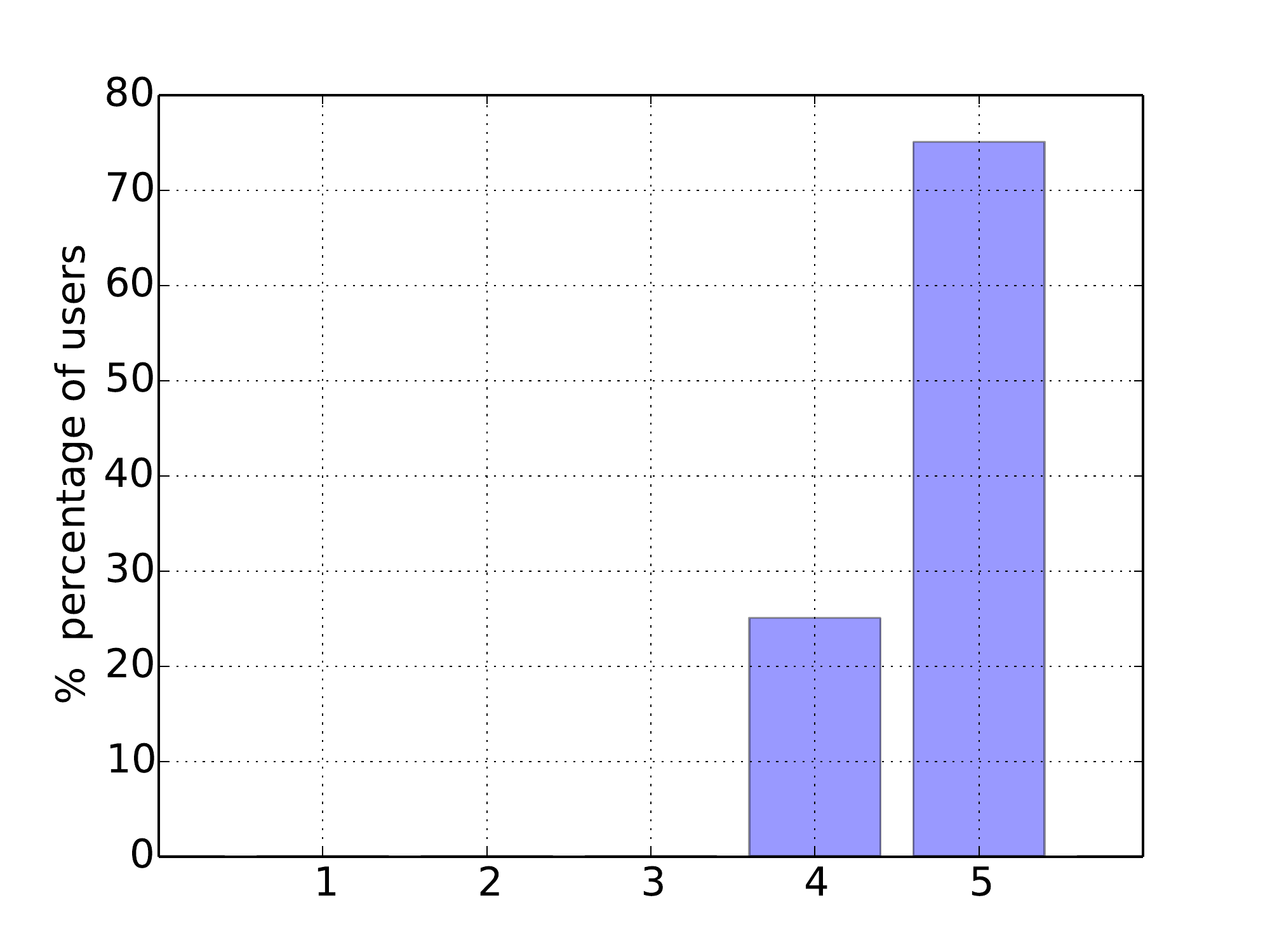}
\vspace{-1em}
\caption{(Custodians) Do you think that CFAS would improve the safety of minors when using OSNs? (1: Totally Disagree, 5: Totally Agree)}
\vspace{-2em}
\label{fig:custodians_safety_improvement}
\end{figure}

\begin{figure}[t!]
\center
\includegraphics[width=0.45\textwidth]{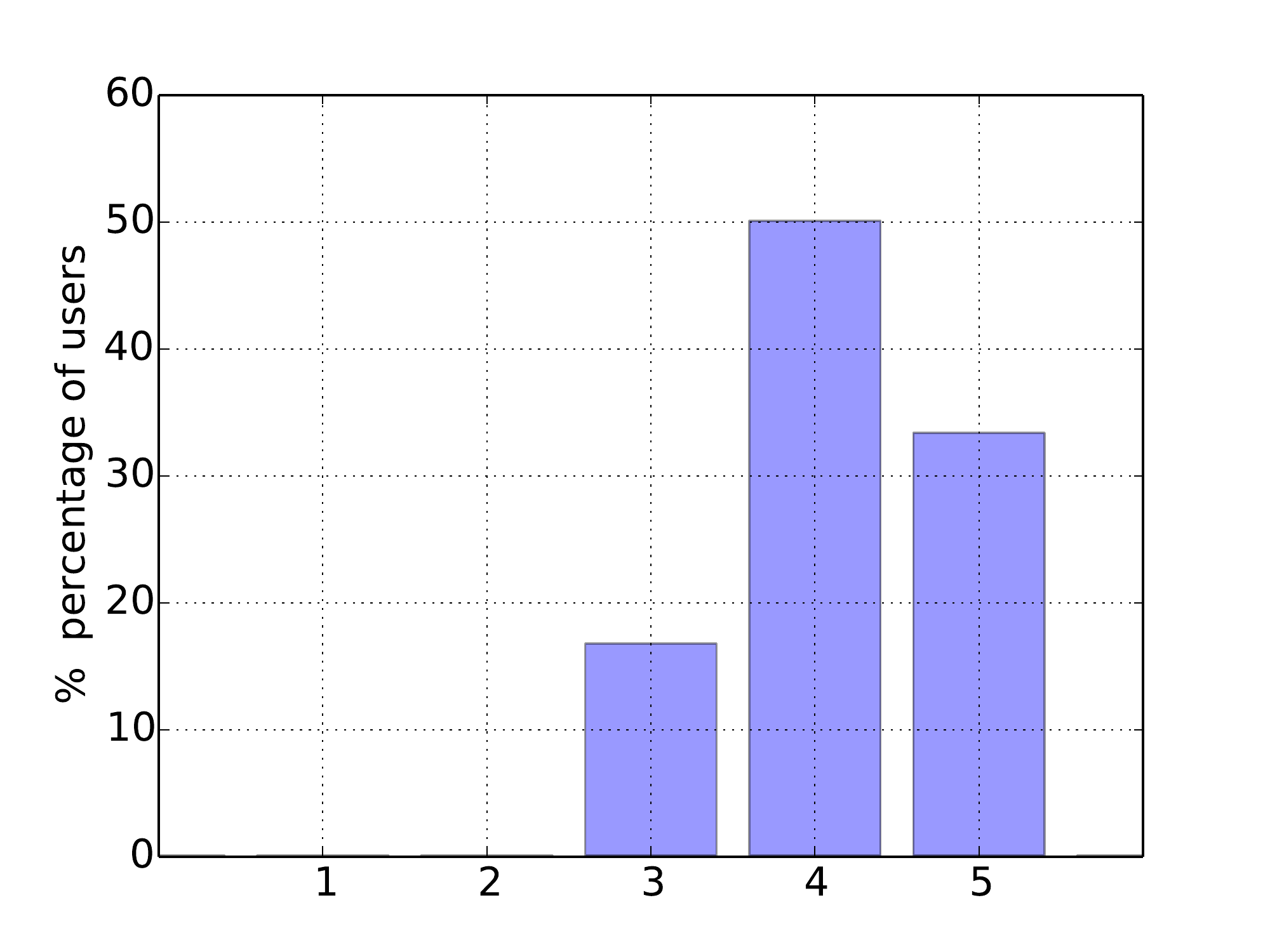}
\vspace{-1em}
\caption{(Custodians) Would you install CFAS at home? (1: Totally Disagree, 5: Totally Agree)}
\vspace{-2em}
\label{fig:custodians_cfas_installation}
\end{figure}

On the other hand, the overwhelming majority of the custodians report that their child never complained of being a victim or a spectator of such threats online (Figure~\ref{fig:custodian_osn_threats}).
Although this is a small number of participants, it depicts that it is usually the case that minors don't report the threats they face on OSNs to their custodians.
Last, all of the custodians agree that CFAS could improve the safety of minors online (Figure~\ref{fig:custodians_safety_improvement}), and the overwhelming majority of custodians report that they would install CFAS at home (Figure~\ref{fig:custodians_cfas_installation}).

\section{Related Work}\label{sec:relatedwork}
This section reviews some web-based and mobile applications that try to protect adolescents on the Internet and OSNs. 
We list the ones most relevant to the concepts of CFAS.

Qustodio is a parental control software \cite{Webpage:2020:Qustodio} that enables parents to monitor and manage their kids' web and offline activity on their devices. 
It also tracks with whom the child communicates on various OSNs and can be used as sensitive content detection and protection tool (using filters).
Last, it monitors messages, calls, and the location of the minor's device. 
Kidlogger allows custodians to monitor what their children are doing on their computer or smartphone \cite{Webpage:2020:KidLogger}. 
It performs keystroke logging, keeps a schedule of which websites the minors visit and what applications they use, and with whom they are communicating on Facebook. 
Also, Kidlogger offers sound recording of phone and online calls, smartphone location tracking, and photo capture monitoring.
Web of Trust (WoT) is a browser add-on and smartphone application for website reputation rating that warns users about whether to trust a website or not \cite{Webpage:2020:WOT}. 

Mspy is a smartphone application that monitors almost all the applications and activities on the smartphone of the minor~\cite{Webpage:2020:mSpy}.
Alarmingly, the application may be installed on the smartphone of the minor by the custodian and remain hidden, so the minor cannot know they are being monitored.
Syfer \cite{Webpage:2020:SYFER} is a device, still in production, that can be plugged into the router of the house network and analyses the traffic activity for possible threats.
It protects against cyber threats in realtime, stops invasive data collection, offers a VPN, has artificial intelligence for enhanced security, and blocks advertisements. 
It doesn't log any information, and it offers encrypted activity. 
It restricts inappropriate content with real-time website analysis provided by their AI engine. 
Bark \cite{Webpage:2020:Bark} monitors text messages, YouTube, emails, and 24 different social networks for potential safety concerns. 
Bark looks for activity that may indicate online predators, adult content, cyberbullying, drug use, suicidal thoughts, and more. 
In case anything suspicious is detected, the custodians receive automatic alerts along with expert recommendations from child psychologists for addressing the issue. 
They offer an application for iOS, Android, Kindle, browser add-ons for Google chrome on PC and Safari on Mac, and Kindle. 
The user has to allow the Bark application to send all the traffic data to Bark's Back-End for analysis and detection.

The majority of the existing applications follows a more traditional approach (monitoring, restrictions over online activities). 
Most applications consider parents or custodians as the end-users, instead of the children \cite{Badillo-Urquiola2019a}\cite{McNally2018}. 
Many of the applications do not have interfaces for children but are just installed as services running in the background \cite{Wisniewski2017}.
A new notion suggests designing and developing tools and software that is more ``children-aware'' and ``children-friendly''. 
Online safety applications should consider the child as the major user and try to enrich children's self-regulation and their risk coping skills in cases of online dangers \cite{Ghosh2020}. 
By enforcing this child-friendly approach, we achieve a collaboration where parents and children need to communicate and discuss online risks and behavior in contrast with the approach of restriction and monitoring. 
We aim to teach children how to cope with online threats and use social media with responsibility and self-awareness. 
CFAS follows this approach by involving the child in the process of setting the filters, and parental and Back-End visibility options.
In addition, the cybersafety tools require the child's consent to be activated.
Last, we note that this work is a follow up of the work presented by Papasavva~\cite{papasavva2019privacy}.

\section{Conclusion}\label{sec:conclusion}
In this paper, we present the architecture of a user-centric privacy-preserving advanced family advice suite for the protection of minors on OSNs. 
The architecture comprises three main components, namely, the Data Analytics Software Stack, the Intelligent Web-Proxy, and a browser add-on, which operates as a guardian angel of the child while using OSNs.
This architecture aims to protect minors when using OSNs while preserving their privacy. 
We propose Guardian Avatars that interact with, warn, and advise adolescences when they face threats on OSNs. 
Also, the custodian of the adolescent receives notifications on their Parental Console in case a malicious activity is detected by the classifiers hosted on the IWP to be aware of the threats their child was exposed to. 
Importantly, the custodian can only see the relevant content, which indicated to be suspicious, only if the minor had previously given their explicit consent. 

Blocking content from the minors or thoroughly monitoring their every online-move should not be the solution as it violates the privacy of the adolescents.  
The proposed architecture advertises the collaboration between parents and children and aims at bringing the family to work together to protect the vulnerable groups of the Internet while using OSNs.

\section*{Acknowledgments}
\small
This project has received funding from the European Union's Horizon 2020 Research and Innovation program under the Marie Skłodowska-Curie ENCASE project (Grant Agreement No. 691025), and the CYberSafety II project (Grant Agreement No. 1614254). 
This work reflects only the authors' views.

\small
\bibliographystyle{IEEEtran}
\bibliography{bibliography}

\end{document}